\documentclass[usenatbib]{mn2e}
\usepackage{epsfig}

\newcommand{\etal}{et~al.} 
\newcommand{\ionhy}{H{\sc ii} }

\newcommand{\degrees}{$^\circ$}
\newcommand{\kms}{$\mbox{km~s}^{-1}$ }
\newcommand{\kmsns}{$\mbox{km~s}^{-1}$}

\newcommand{\water}{H$_{2}$O }
\newcommand{\CO}{$^{13}$CO }
\newcommand{\COns}{$^{13}$CO}

\newcommand{\vsfig}[2]           % Single FIGure (put one figures in the
{
  \begin{center}
    \begin{minipage}[t]{0.05\textwidth}
      {\footnotesize \raisebox{40mm}{(#2)}}
    \end{minipage}
    \begin{minipage}[t]{0.42\textwidth}
      \epsfig{file=./#1.ps,height=0.95\textwidth,angle=270}
    \end{minipage}
    \hfill
  \end{center}
}

\newcommand{\specdfig}[2]        % Double FIGures (put two figures  
{
   \begin{center}
     \begin{minipage}[t]{0.45\textwidth}
         \epsfig{file=eps/#1.eps,height=0.8\textwidth,angle=270}
     \end{minipage}
     \hfill
     \begin{minipage}[t]{0.45\textwidth}
         \epsfig{file=eps/#2.eps,height=0.8\textwidth,angle=270}
     \end{minipage}
   \end{center}
}

\newcommand{\specsfig}[1]        % Single FIGure (put one figure of  
{
   \begin{center}
     \begin{minipage}[t]{0.45\textwidth}
         \epsfig{file=eps/#1.eps,height=0.8\textwidth,angle=270}
     \end{minipage}
   \end{center}
}

\newcommand{\boxfig}[1]        % Single FIGure (put one figure of  
{
   \begin{center}
     \begin{minipage}[t]{0.46\textwidth}
         \epsfig{file=eps/#1.eps,height=0.45\textwidth,angle=270}
     \end{minipage}
   \end{center}
}

\newcommand{\twofig}[2]        % Double FIGures (put two figures  
{
   \begin{center}
     \begin{minipage}[t]{0.5\textwidth}
         \epsfig{file=eps/#1.eps,height=0.95\textwidth}
     \end{minipage}
     \hfill
     \begin{minipage}[t]{0.5\textwidth}
         \epsfig{file=eps/#2.eps,height=0.95\textwidth}
     \end{minipage}
   \end{center}
}

\begin{document}

\title[A search for 22-GHz water masers] {A search for 22-GHz water masers
  within the giant molecular cloud associated with RCW 106}
\author[S.\ L. Breen \etal]{S.\ L. Breen,$^1$  S.\ P. Ellingsen,$^1$ M. Johnston-Hollitt,$^1$ S. Wotherspoon,$^{1}$   I. Bains,$^{2,3}$           \newauthor M.\ G. Burton,$^2$ M. Cunningham,$^2$ N. Lo,$^2$ C.\ E. Senkbeil$^1$ and T. Wong$^{2,4}$\\
  \\
  $^1$ School of Mathematics and Physics, University of Tasmania, Private Bag 37, Hobart, Tasmania 7001, Australia;\\
  Shari.Breen@utas.edu.au\\
  $^2$ School of Physics, University of New South Wales, Sydney, NSW 2052, Australia\\
  $^3$ Centre for Astrophysics and Supercomputing, Swinburne
  University of Technology, PO Box 218, Hawthorn, VIC 3122,
  Australia\\
  $^4$ Australia Telescope National Facility, CSIRO, PO Box 76, Epping, NSW 1710, Australia}

 \maketitle
  
 \begin{abstract}
  
   We report the results of a blind search for 22-GHz water masers in
   two regions, covering approximately half a square degree, within
   the giant molecular cloud associated with RCW 106. The complete
   search of the two regions was carried out with the 26-m Mount
   Pleasant radio telescope and resulted in the detection of nine
   water masers, five of which are new detections.  Australia Telescope
   Compact Array (ATCA) observations of these detections have allowed
   us to obtain positions with arcsecond accuracy, allowing meaningful
   comparison with infrared and molecular data for the region.  We find
   that for the regions surveyed there are more water masers than
   either 6.7-GHz methanol, or main-line OH masers.  The water masers
   are concentrated towards the central axis of the star formation
   region, in contrast to the 6.7-GHz methanol masers which tend to be
   located near the periphery.  The colours of the GLIMPSE point
   sources associated with the water masers are similar to those of
   6.7-GHz methanol masers, but slightly less red.  We have made a
   statistical investigation of the properties of the \CO and 1.2-mm
   dust clumps with and without associated water masers.  We find that
   the water masers are associated with the more massive, denser and
   brighter \CO and 1.2-mm dust clumps.  We present
   statistical models that are able to predict those \CO and 1.2-mm dust clumps that are
   likely to have associated water masers, with a low misclassification rate.
  
\end{abstract}

\begin{keywords}
 masers - ISM: molecules - radio lines: general - stars: formation.
\end{keywords}

\section{Introduction}

Water masers have been regarded as one of the best indicators of star 
formation since soon after the first detection of 22-GHz water maser emission in
the sources Sgr B2, Orion and W 49 in 1969 \citep{Cheung69}.
The 22-GHz 6$_{1,6}$ $\rightarrow$ 5$_{2,3}$ rotational transition of
\water is the brightest spectral line at radio wavelengths and traces
shocked gas in star formation regions, outflows, as well as dense
circumstellar shells around evolved stars.  Emission from this
transition often exhibits significantly greater temporal variability
than is commonly observed in other interstellar maser species such as
OH and methanol \citep{Brand2003}. The physical conditions required to
produce water maser excitation are high densities (10$^7$-10$^9$
cm$^{-3}$) and temperatures of a few 100 K \citep*{Elitzur1989}, both
of which are seen in the inner parts of circumstellar disks around
young stellar objects and in regions of shocked gas
\citep{Torrelles2002}.

Interstellar masers from water \citep[e.g.][]{Valdettaro2001}, OH
\citep[e.g.][]{C98} and methanol \citep*[e.g.][]{Pestalozzi2005}
transitions have been detected towards hundreds of star formation
regions in our Galaxy, with many of these showing emission from more
than one species.  While there have been a number of large-scale
untargeted searches for OH \citep{Caswell} and methanol
\citep{Simon96,Szymczak2002}, previous searches for water masers in
star formation regions have typically targeted ultracompact \ionhy
regions selected on the basis of {\em IRAS} (Infrared Astronomy Satellite) colours
\citep*[e.g.][]{Churchwell1990}, or other sources believed to be
high-mass young stellar objects \citep*[e.g.][]{Beuther2002}.  To date
there have been no large untargeted surveys for water masers
primarily because at a frequency of 22~GHz telescope beam sizes are
approximately one-third the size of those at 6.7-GHz, the frequency of the
strongest methanol maser transition, and hence require approximately
an order of magnitude more pointings.  Here we present an untargeted
search for water masers within the giant molecular cloud (GMC) complex
associated with RCW 106 (the G\,333.2--0.6 GMC). The GMC is located at a distance of 3.6 kpc
\citep{Lock79} and was discovered by \citet{G77} during observations
of molecular clouds associated with southern Galactic \ionhy regions
in the J=1--0 transition of CO. These observations uncovered a number
of bright \ionhy regions along a line which is almost parallel to the
Galactic plane including one of the brightest infrared sources in our
Galaxy, {\em IRAS}\,16183--4958 \citep{Beck}, which is associated with the
\ionhy region G\,333.6--0.2.

The GMC is roughly centred on $l$ $\sim$ 333\degrees, $b$ $\sim$
\mbox{--0\fdg5} (or $ \alpha_{j2000}$=16:21, $\delta_{j2000}$=--50:20) and
extends approximately 1\fdg2 x 0\fdg6 on the sky (or approximately 90
pc x 30 pc at an assumed distance of 3.6 kpc \citep{B06}). This region
passes through the ring of molecular clouds that circle the Galaxy at
around 5 kpc from its centre \citep[e.g.][]{Simon2001} and exhibits a
diverse range of molecular regions, bright \ionhy regions, GLIMPSE
({\em Spitzer} Galactic Legacy Infrared Mid-Plane Survey
Extraordinaire) point sources, {\em IRAS} and {\em MSX} (Midcourse
Space Experiment) sources, all of which are embedded in a larger region
of diffuse atomic and molecular gas.

Observations of the \CO J=1--0 transition at 110 GHz by \citet{B06}
showed the velocity structure of the region to contain five primary
velocity components, with the dominent feature centred on $v_{\rm LSR}$
$\sim$ --50 \kmsns. \citet{B06} showed that it is likely that at least five 
distinct molecular clouds lie along the line-of-sight (see Section~\ref{section:foreground_cloud}).
For this reason the G\,333.2--0.6 GMC is refered to as the main cloud, showing emission 
over the velocity range -35 \kms to -65 \kmsns.  Analysis of the integrated \CO data using the
{\scshape{clumpfind}} algorithm of \citet{Williams94}
identified 61 \CO clumps within the main cloud. The \CO emission takes the form of a string
of knots with the clumps arranged along an axis aligned NW to SE
\citep{B06}.

This GMC has been the focus of numerous observations in recent times,
including far-infrared (FIR) observations of the dust continuum at 150
and 210 $\mu$m which identified 23 emission peaks with dust
temperatures between 20 and 40 K \citep{Karnik01}. The region was also
observed by \citet{Mook04} who used SIMBA (SEST IMaging Bolometer Array)
on the Swedish European Southern Observatory Submillimetre Telescope
(SEST) to obtained a 1.2-mm dust continuum image of the region.
\citet{Mook04} identified 95 dust emission peaks (or clumps), half of
which have {\em MSX} counterparts.  Observations of a multitude of
molecular lines towards detected \ionhy regions and {\em IRAS} sources
indicate probable ongoing star formation \citep{Mook04}. These
observations, like those of \citet{B06} have identified that the GMC
has a linear clumpy structure.

Complete surveys of the GMC region have been carried out by
\citet{Simon96} for 6.7-GHz methanol masers and \citet{Caswell} for
1665- and 1667-MHz OH masers. These surveys resulted in the detection
of nine methanol and six OH masers within the region surveyed in \CO
by \citet{B06}. Five water masers have also been detected within this
region in targeted searches made by \citet{J72}, \citet{C74},
\citet{B80} and \citet{BS82}.

\begin{figure}
	\epsfig{figure=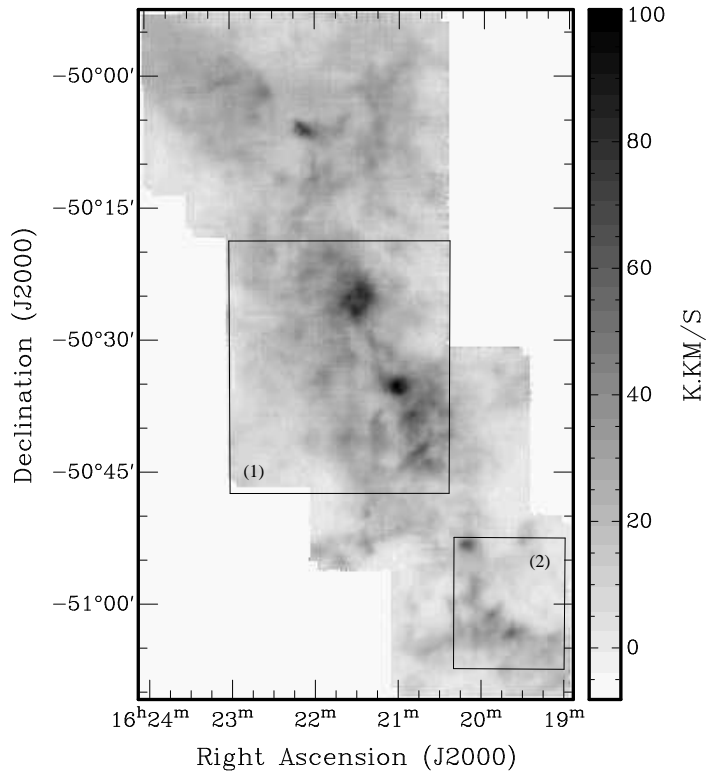,width=10cm}
	\caption{Integrated \CO emission observed by \citet{B06} with
          the two regions surveyed for water masers overlaid.}
  \label{fig:13COregions}
\end{figure}

Here, we present the results of an untargeted survey of 22-GHz water masers made with the
University of Tasmania Mount Pleasant 26-m radio telescope.  The survey 
covers two distinct regions within the GMC.
The first (hereafter Region 1) is a 0\fdg50 x 0\fdg42 area centered on
$l$ $\sim$ 333, $b$ $\sim$ --0\fdg5. This region encompasses much of
the high density gas and dust regions identified by \citet{Karnik01},
\citet{Mook04} and \citet{B06} in the central section of the
GMC. Region 1 contains one previously detected 22-GHz water
maser, G\,333.13--0.43, discovered by \citet{C74}. The second region
(hereafter Region 2) covers a 0\fdg28 x 0\fdg24 area of the GMC and
is approximately centred on well-known optically visible H{\sc ii}
region RCW 106. This region contains two previously detected water
masers. The extent of the two regions compared to the integrated \CO
emission are shown in Fig.~\ref{fig:13COregions}.

The GMC is the focus of an ongoing project to characterise the
turbulence in the molecular cloud and compare this to the star
formation efficiency in order to attain a relationship between the
two.  Commencing in 2004 a multitude of millimetre molecular line
transitions (including \CO \citep{B06}, CS, C$^{34}$S, C$^{18}$O,
C$_{2}$H, HCN, H$^{13}$CN, HCO$^{+}$, H$^{13}$CO$^{+}$, HNC,
CH$_{3}$OH and SiO) have been observed by the Delta Quadrant Survey
Team at the University of New South
Wales\footnote{http://www.phys.unsw.edu.au/astro/mopra/dqs.php} (UNSW).
Interstellar masers require special physical conditions and the
different species are generally thought to trace particular
evolutionary phases of the high-mass star formation process.  Through
combining information on all the strong and common interstellar maser
transitions with millimetre molecular line and other existing infrared
and millimetre continuum data we aim to better understand the
evolutionary phases traced by each type of maser.

\section{Observations \& Data processing}

The primary search for 22-GHz water masers towards the G\,333.2--0.6
GMC was undertaken between 2005 April-November using the University of
Tasmania's 26-m radio telescope at Mount Pleasant. The observations
were made with a cryogenically cooled receiver that detects both left
and right circularly polarized signals and has a typical system
equivalent flux density of 2000-2200~Jy in good weather at elevations
above about 40~degrees.  At 22~GHz the telescope has a 2.2-arcmin
half power beam width (HPBW) and at the time of the observations the
measured RMS pointing errors were $\sim$1 arcminute.  The data were
recorded using a 2-bit auto-correlation spectrometer configured with
2048 spectral channels per polarisation over a 32-MHz bandwidth, which
was centred on an LSR velocity of approximately \mbox{--40~\kmsns}. The
observations covered a velocity range of 430 \kms with a spectral
resolution of 0.25 \kmsns.  The two regions were surveyed using an
equilateral triangle pattern with each pointing separated by 1.1
arcminutes (half of the HPBW) from all adjacent grid points. Each
pointing was observed for a total onsource integration time of 10
minutes. The water maser G\,333.608--0.215,
discovered by \citet{J72}, was observed at the beginning of each
observing session to test the system and ensure consistency.  Although
it is not located in either of the survey regions we have included it
in the nine masers detected in this survey.

The weather conditions in which the observations were made varied
substantially, and in general, the data for Region 1 were taken 
under much more favourable conditions. In order to
minimize atmospheric and telescope pointing effects all observations
were made above an elevation of 40 degrees.  Data affected by poor
weather conditions were reobserved (in some cases on multiple
occasions), however a small percentage of the data has a significantly
poorer sensitivity.  The observations of Region 1
required approximately 750 pointings while Region 2 was much smaller
containing just under 300 pointings. After box car smoothing over
5 channels and averaging the two polarisations together the majority (78 percent) of the
data taken in Region 1 had an RMS noise level of less than 1 Jy while
the remaining (22 percent of) data in this region was subject to an
RMS noise level of between 1 Jy and 2 Jy. This equates
to a 5$\sigma$ detection limit of 5 Jy over the majority of the
region, with a maximum of 10 Jy in the worst affected pointings. In
Region 2 only 35 percent of pointings have an RMS noise level of less
than 1 Jy and the remaining 65 percent had an RMS noise of up to 5
Jy. This equates to a 5$\sigma$ detection limit between 5 and 25 Jy,
hence our ability to detect weaker masers within Region 2 is greatly
reduced.

For each detected water maser additional observations consisting of a
5-point grid (centred on the preliminary position) were made to locate the emission 
more accurately.  The position was determined
by fitting a 2D circular Gaussian (with the same HPBW as the
telescope) to the relative amplitudes of the strong emission in the
maser spectra observed in the 5-point grid.  Positions determined in
this manner are accurate to approximately 1 arcminute, which is
insufficient to allow meaningful comparison with millimetre and
infrared observations of the region.  In order to obtain accurate
positions for the detected water masers we were granted two sessions
of Australia Telescope Compact Array (ATCA) director's
time. Preliminary observations of the masers detected in the Mount
Pleasant survey were made on 2006 June 19 with the ATCA in the 1.5D
configuration.  The observations were centred on a frequency of 22.238
GHz with the correlator configured to record 256 spectral channels
across a 16-MHz bandwidth. These observations failed to detect one of
the water masers discovered in the Mount Pleasant survey
(G\,333.29--0.38), most likely due to temporal variability.

Further observations were made with the ATCA in the 6A configuration
on 2006 July 16 \& 17.  In this array configuration the minimum
baseline length is 337 m and the maximum is 5939 m. The observations
were centred on 22.236 GHz and the correlator sampled two orthogonal
linear polarisations, each processed to give a 512 channel spectrum
across an 8-MHz bandwidth. Each of the maser sites detected in the
earlier ATCA observations (8 in total) were observed in a series of
three minute cuts over a range of hour-angles.  Observations of one of
the phase calibrators MRC\,B1613--586 or MRC\,B1646--50 were made for 
a duration
of 90 seconds before and after every two maser observations (i.e. every
6 minutes). PKS\,B1934--638 was used as the primary flux calibrator and
at 22.236 GHz has an assumed flux density of 0.83~Jy. PKS\,B1253--055
was used for bandpass calibration. These ATCA observations were made
using reference pointing, and as arcsecond accurate positions were
obtained in the initial ATCA observations all of the maser sites were
close to the centre of the primary beam. Over the two days each source
was observed a total of eight times, equivalent to a total on-source
integration time of 24 minutes.

The data were processed using the {\tt miriad} software package \citep{Sault}
applying the standard techniques for ATCA spectral line and continuum
observations.   Continuum sources which were located
away from the centre of the primary beam have had their flux densities
corrected to account for beam attenuation.  The frequency resolution, after Hanning smoothing, was
0.038 MHz or 0.50 \kmsns. The RMS noise in a single spectral
channel in the final data cubes was approximately 0.15 Jy and the
signal-to-noise ratio of the final spectra was about 20:1 in the worst
case.  The RMS noise level in the continuum images was typically around 0.02 Jy/beam.  These observations have enabled us to determine the positions
of the water masers to an accuracy of approximately 0.5~arcseconds.

\section{Results}

\begin{figure*}
\vspace{-2cm}
  \epsfig{file=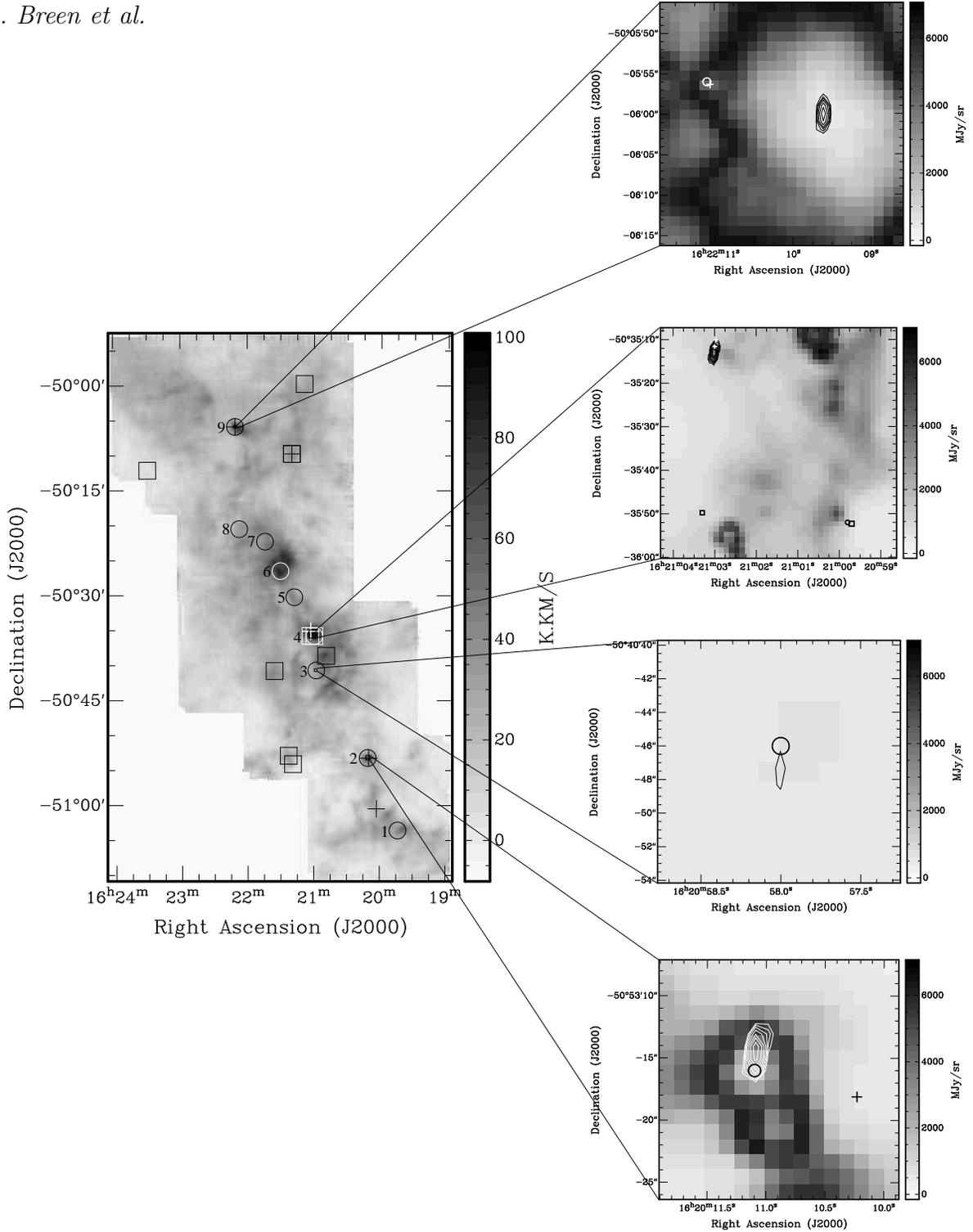,width=17cm}
  \caption{The main image shows the integrated \CO emission observed
    by \citet{B06} with the positions of water (circle), methanol
    (square) and OH (cross) masers overlaid (note that the most central OH maser in the main image is in fact two OH sources close together as shown in the second sub-image).  Also overlaid are the water maser source numbers as shown in Table~\ref{tab:masers}. The positions of the
    methanol and OH masers have also been obtained from ATCA
    observations and have a similar positional accuracy to the water
    maser positions \citep{C98,Simon05}.  In the main image, the size
    of the shapes is much larger than the positional accuracy,
    however, the four sub-images (of water masers 9, 4, 3 and 2 respectively) show the positions of the maser
    species and indicate the positional accuracy of these masers
    overlaid on a 8.0-$\mu$m GLIMPSE image of the region. Also
    overlaid on the sub-images are the 22-GHz continuum contours
    detected in our ATCA observation.  The first contour in each case
    is at the 5$\sigma$ level for the continuum image and they
    increase in factors of $\sqrt{2}$. Details of the continuum
    sources can be found in Table~\ref{tab:continuum}.}
  \label{fig:sick_pic}
\end{figure*}

\begin{figure*}
  \vspace{-2cm}
  \epsfig{file=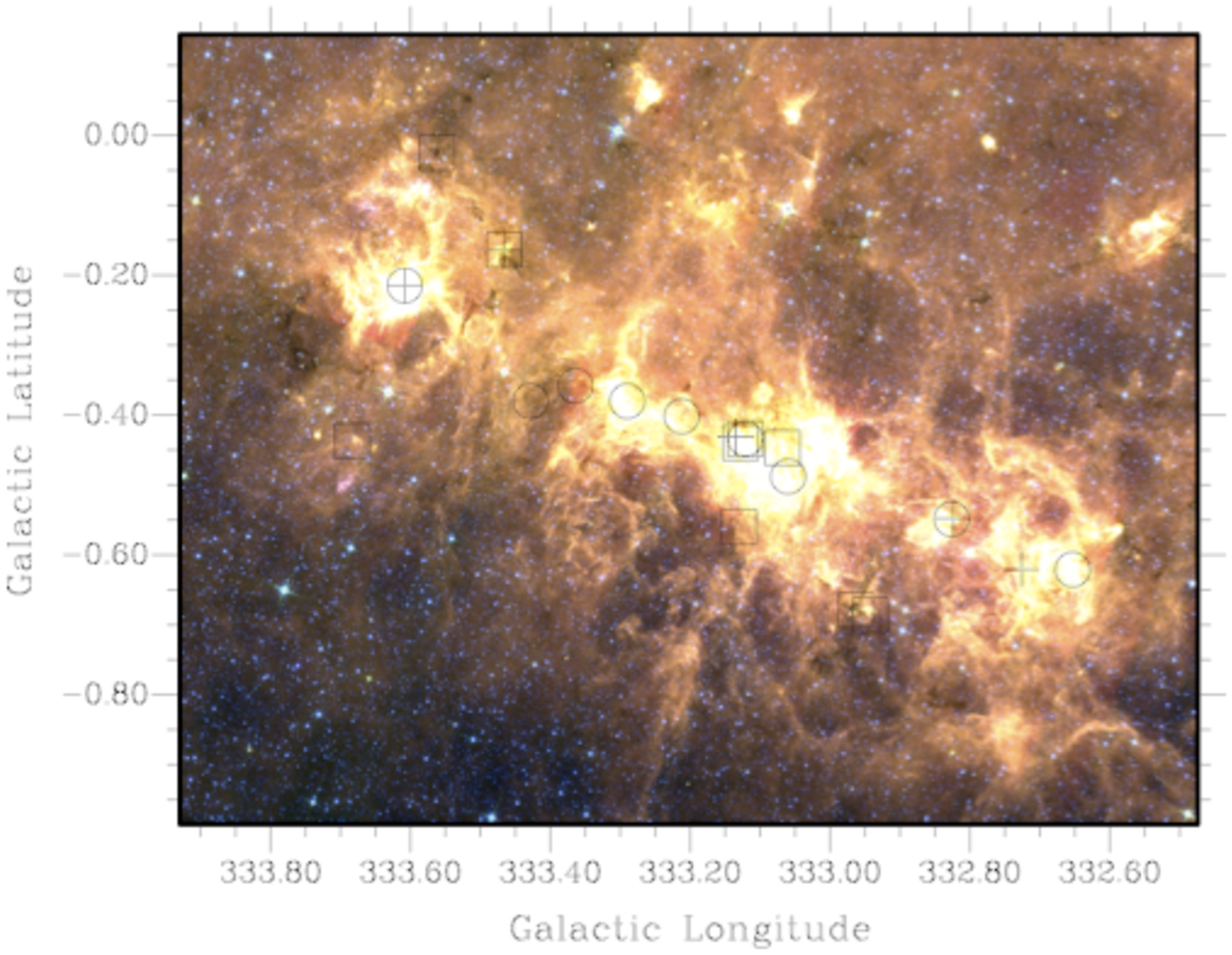,width=18cm}
  \vspace{-1cm}
  \caption{Three colour GLIMPSE image of the GMC where red=8.0-$\mu$m
    green=5.8-$\mu$m and blue=3.6-$\mu$m. The positions of the nine
    water masers detected in this survey are represented by circles,
    positions of the methanol masers observed by \citet{Simon05} are
    represented by squares and the positions of the OH masers observed
    by \citet{C98} are represented by crosses. Water maser sources 1-9 are numbered in order of increasing galactic longitude.}
  \label{fig:3colour}
\end{figure*}

A search of two regions near the G\,333.2--0.6 giant molecular cloud
resulted in the detection of nine 22-GHz water masers, five of which
are new detections (Table~\ref{tab:masers}), as well four 22-GHz
continuum sources (see Section~\ref{section:cont}).
Figure~\ref{fig:sick_pic} shows the positions of all the detected
water masers overlaying the integrated \CO emission observed by
\citet{B06}, while Fig.~\ref{fig:3colour} shows the maser locations
on a three colour GLIMPSE image of the GMC. Comparison of the water maser locations with the other maser species,
the integrated \CO emission and the three colour GLIMPSE image shows
that in general the water masers originate very close to the higher
density molecular gas and warm dust, near the main axis of star
formation within the molecular cloud.  In contrast the methanol masers
tend to be offset from this axis, close to the interface between the intense
mid-infrared emission and the larger molecular cloud
\citep{Ellingsen2006}.

\begin{figure*}
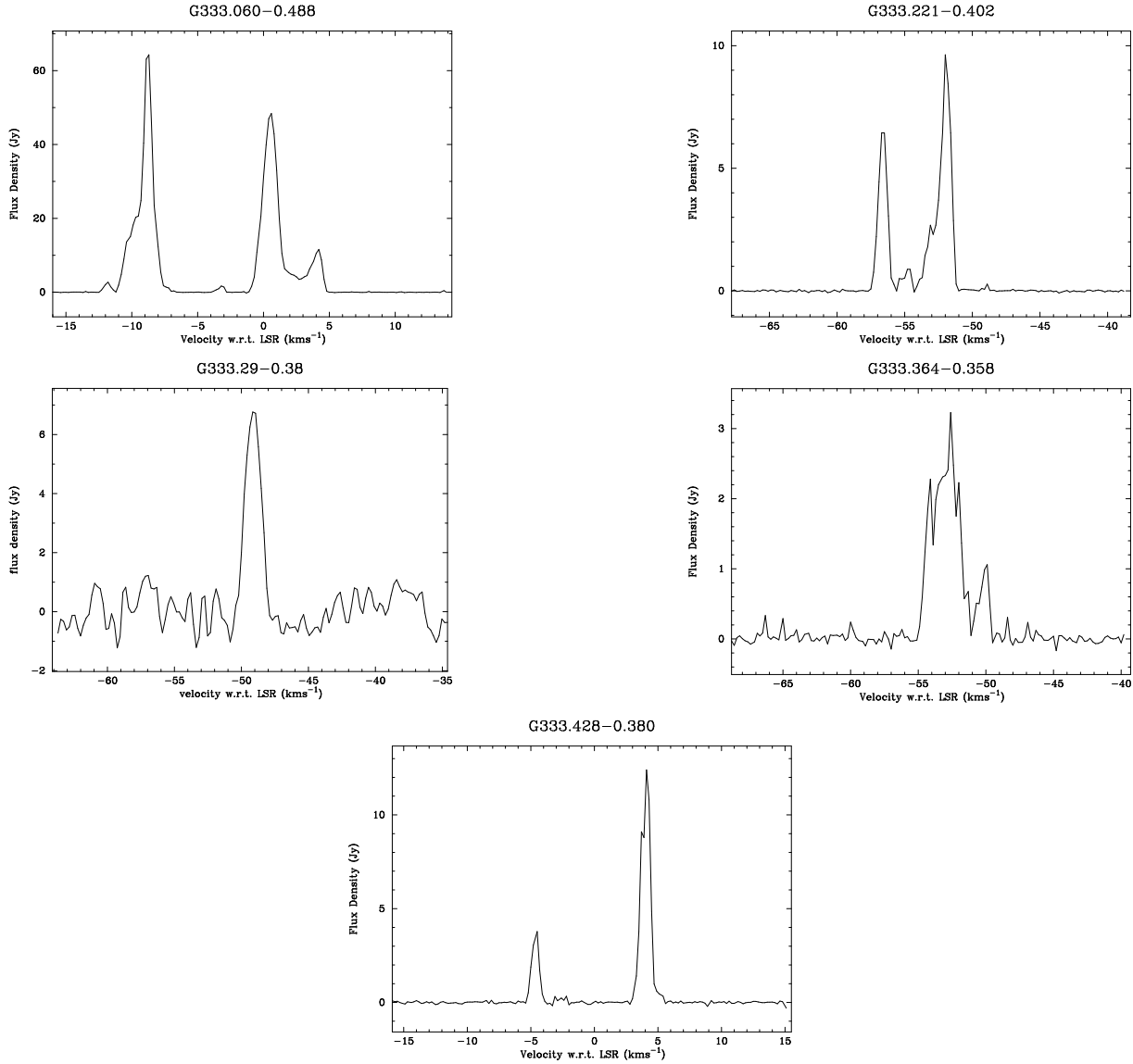

	\specdfig{fig4_1}{fig4_2}
	\specdfig{fig4_3}{fig4_4}
	\specsfig{fig4_5}
	\caption{Spectra of the 22-GHz water maser sources discovered
          in this survey. All of the spectra presented are from the
          ATCA observations with the exception of G\,333.29--0.38 which is
          from the Mount Pleasant observations.}
	\label{fig:new}
\end{figure*}

\begin{figure*}
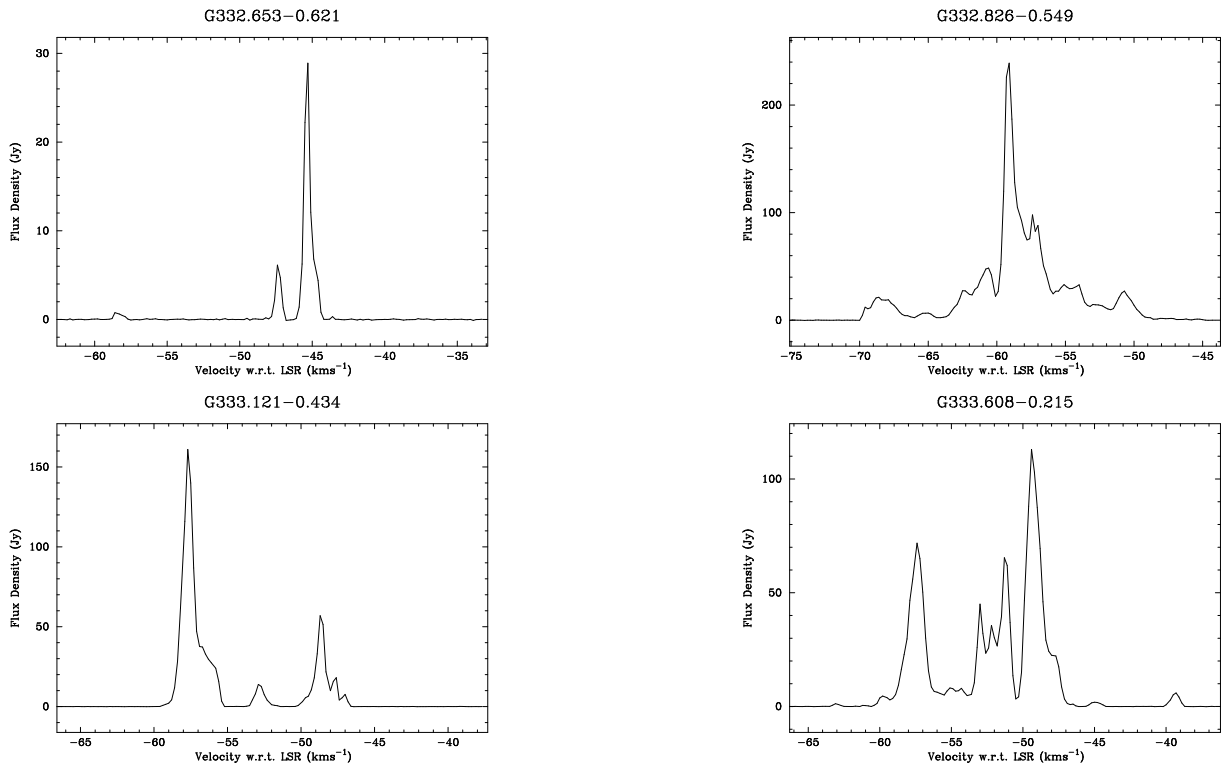

	\specdfig{fig5_1}{fig5_2}
	\specdfig{fig5_3}{fig5_4}
	\caption{ATCA spectra of the 22-GHz water maser sources
          detected in the search that have been previously
          discovered.}
	\label{fig:previous}
\end{figure*}

Spectra of the detected water masers are shown in
Figs.~\ref{fig:new} and~\ref{fig:previous}.  The spectra have been
produced by integrating the emission in the ATCA image cubes for each
source.  The only exception is water maser G\,333.29--0.38 for which
the Mount Pleasant spectrum is shown.  The positional accuracy of
sources detected in the ATCA observations is approximately 0.5
arcseconds and we have used three significant figures after the
decimal place in their Galactic coordinate names.  For the source
only detected in the Mount Pleasant component of the survey we are
only justified in using 2 significant figures and have done so
throughout the paper.  Comments on each maser can be found in
section~\ref{section:sources}.  The newly detected water masers
(Fig.~\ref{fig:new}), with one exception have a peak flux density less
than 50~Jy, while the previously detected sources
(Fig.~\ref{fig:previous}), again with one exception, have peak flux
densities greater than 100~Jy.

The 6.7-GHz methanol masers sites in this region have previously been
searched for associated 22-GHz water maser emission by
\citet{Hanslow1997} who detected emission towards a number of sources
(G\,332.942--0.686, G\,333.121--0.434, G\,333.128--0.440,
G\,333.130--0.560, G\,333.234--0.062 and G\,333.466--0.164) in the
G\,333.2--0.6 giant molecular cloud.  Three of these (G\,332.942--0.686,
G\,333.234--0.062 and G\,333.466--0.164) lie outside the regions of our
untargeted search.  While of the other three, only G\,333.121--0.434
was detected in the current work.  It appears that the emission
attributed to G\,333.128--0.440 by \citet{Hanslow1997} is in fact
associated with G\,333.121--0.434, while that associated with
G\,333.130--0.560 was not detectable at the epoch of our search.
Because of the uncertainty in the position of the water masers
detected by \citet{Hanslow1997} and the possibility that some may
result from an unassociated strong source detected in a sidelobe, in
Figs.~\ref{fig:sick_pic} \& \ref{fig:3colour} we have only marked
those water masers detected in our current survey.

In addition to the nine water masers that we present, we made a
one-time detection with the 26-m Mount Pleasant radio telescope of a
water maser right on the edge of the field of observations with
coordinates G\,333.22--0.20. Subsequent observations showed no
detectable emission and as a result we do not include it in the list
of water masers that we detect. This emission consisted of a single
velocity feature at --87 \kms of around 6 Jy. We believe that this
emission was actually a detection of the strong water maser
G\,333.234--0.062 detected by \citet{Hanslow1997}. \citet{Hanslow1997}
reported G\,333.234--0.062 to consists of multiple velocity features
with the most prominent observed at --86 \kmsns, with a flux density of
108 Jy.

The majority of the water masers we detected have exhibited
variability of up to a factor of 10 on a time scale of several months.
This type of variability is common in water masers and a survey of
water maser emission towards main-line OH masers in star formation
regions by \citet{B80} found that about 60 percent of water masers
exhibited variability of up to a factor of two over an eight month
period, while the remaining 40 percent exhibited more extensive
variability.  Given that our initial observations were made in
varying weather conditions with comparitively poor pointing accuracy
it is difficult to quantify the absolute
variations accurately. However, because the water masers have multiple spectral
components we are able to determine that variability has occurred by
examining the relative amplitudes. Water masers associated with low
mass stars are typically both weaker and more variable than those
associated with high mass stars \citep{Claussen1996}.  So we would
expect our observations to be more likely to detect masers associated
with high-mass star formation, than those associated with less
luminous objects.  Variability is also a likely explanation as to why
two of the masers detected in the survey with the 26-m Mount Pleasant
radio telescope were not detected in the final observations made with
the ATCA 6-8 months later, particularly if these masers are associated
with low-mass stars.

We have compared the positions of the water masers that we detected
with GLIMPSE, {\em IRAS} and {\em MSX} sources, as well as 1.2-mm dust
clumps \citep{Mook04}, FIR sources \citep{Karnik01}, \CO emission
\citep{B06}, CS emission (N. Lo, private communication) and other
maser species.  The relative positional accuracy of each of these
datasets differs, some having significantly poorer positional accuracy
than our ATCA observations.  We consider a water maser to be
associated with a GLIMPSE, {\em IRAS}, {\em MSX} or FIR source if it
is within a radius of 2, 30, 5 or 60 arcseconds respectively.  For
water maser G\,333.29--0.38 we use more relaxed positional constraints
as its position is much less accurately known. For this source we
consider the water maser to be associated if it is within 5, 90, 40 or
60 arcseconds of a GLIMPSE, {\em IRAS}, {\em MSX} or FIR source
respectively.  We consider a water maser to be associated with a \COns,
CS or 1.2-mm dust continuum clump if its position falls within the
radius of the \COns, CS or 1.2 mm dust clumps. We have used the clump radii reported by \citet{B06} and \citet{Mook04} for the \CO and 1.2-mm dust clumps, respectively. In the case of the CS data we have considered that 
a water maser falls within the radius of a given CS clump if it lies within emission that has a value of flux
density per beam which surpasses a 5$\sigma$ level.
Velocity ranges of the \CO and CS clumps 
correlate well with the velocity ranges of the associated masers (see Tables~\ref{tab:CO} and \ref{tab:CS}). 
Table~\ref{tab:ass} summarises all associations.

Of the nine water masers that we detect, four are associated with a
GLIMPSE point source, three are associated with an {\em IRAS} source,
two are associated with an {\em MSX} source, five are associated with
a FIR clump \citep{Karnik01} and seven are associated with a 1.2-mm
dust clump \citep{Mook04}. All of the water masers detected in this
survey either lie within a \CO clump identified by \citet{B06} or an
identifiable emission peak of the \CO data.  In addition all of the
water masers, for which CS data of the GMC was available, are
associated with CS emission peaks covering a comparable velocity range
to the masers.

\begin{table*}
  \caption{22-GHz water masers detected within the survey regions. Column 1 gives the water maser source number (which is used in later tables), column 9 indicates whether or not each maser was detected in the ATCA observation and column 10 gives the water maser references. ATCA positions are quoted for all water masers with the exception of G\,333.29-0.38 (source number 6). References: * = new source; 1 = \citet{C74};  2 = \citet{K76};  3 = \citet{BS82} 4 = \citet{J72}.}
  \begin{tabular}{lcccrrcccl} \hline
    \multicolumn{1}{c}{\bf Source} & {\bf Water} & {\bf Right} & {\bf Declination} & 
      \multicolumn{1}{c}{\bf Peak} & {\bf Peak Vel.}  & 
      \multicolumn{1}{c}{\bf Velocity} & {\bf epoch} & {\bf ATCA} &{\bf Ref} \\
   \multicolumn{1}{c}{\bf number} & {\bf maser}  & {\bf Ascension}         & {\bf (J2000)}     & 
      \multicolumn{1}{c}{\bf Flux} & {\bf wrt LSR}  &
      \multicolumn{1}{c}{\bf Range} &   &{\bf detection?} \\	
       & {\bf  ($l,b$) }          & {\bf (J2000)}                  &             &
      \multicolumn{1}{c}{\bf (Jy)} & {\bf (\kmsns)}&{\bf (\kmsns)}& \\ \hline
     1 & G\,332.653--0.621 & 16:19:43.569 & --51:03:37.06 &  28.9 & --45.3 & --59,--43 & 2006 July &yes &  2\\ 
     2 & G\,332.826--0.549 & 16:20:11.089 & --50:53:16.07&  239.4 & --59.1 & --69,--45 &2006 July & yes & 3 \\ 
     3 & G\,333.060--0.488 & 16:20:58.002 & --50:40:46.32 & 64.3 &--8.7  & --13,5 & 2006 July & yes &*  \\ 
     4 & G\,333.121--0.434 & 16:20:59.762 & --50:35:51.55 & 161.1 &--57.7  &--60,--46  &  2006 July & yes &1\\ 
     5 & G\,333.221--0.402 & 16:21:17.913 & --50:30:17.99 & 9.6 & --52.0 & --58,--48 & 2006 July &yes &*\\     
     6 & G\,333.29--0.38 & 16:21:30.4 & --50:26:34 & 7 & --49.0 & --51,--47 & 2005 June & no & *\\
     7 & G\,333.364--0.358 & 16:21:44.319 & --50:22:21.08 & 3.2 & --52.6 & --55,--49 & 2006 July & yes &*\\ 
     8 & G\,333.428--0.380 & 16:22:07.539 & --50:20:34.29 &  12.4& 4.1 &--6,6& 2006 July & yes &*\\  
     9 & G\,333.608--0.215 & 16:22:11.060 & --50:05:55.98 & 103.3 & --49.2 &--64,--38  & 2006 July &yes &4\\\hline
       \end{tabular}
  \label{tab:masers}
\end{table*}

\subsection{Individual sources}\label{section:sources}

Comments on the nine water masers that we detect can be found in Section~\ref{sect:comments_sources}. Refer to Table~\ref{tab:masers} for a summary of each of the sources. Figures~\ref{fig:sick_pic} and \ref{fig:3colour} show the positions of the water masers overlaying the integrated \CO emission \citep{B06} and a three colour GLIMPSE image of the GMC (also showing the locations of 6.7-GHz methanol masers \citep{Simon05} and main-line OH masers \citep{C98}), respectively.

\subsubsection{Possible associations with a foreground cloud}\label{section:foreground_cloud}

\citet{B06} found that the mean velocity profile of the \CO emission averaged 
over their field of observations exhibited five distinct velocity features 
(see Fig.~\ref{fig:velocity_profile}), while only the brightest velocity 
feature was attributed to the G\,333.2--0.6 GMC. This most prominent feature 
is also the broadest feature and spans a velocity range of --35 \kms 
to --65 \kmsns. The additional four velocity features 
seen in Fig.~\ref{fig:velocity_profile} are centred on velocities of 
--105, --90, --70 and --10 \kmsns. \citet{B06} plotted 
the kinematic distance versus the LSR velocity at the centre of the GMC and 
found that for the clouds at --10, --50, --70, --90, --105 \kms the 
associated distances were approximately 1, 3.5, 4.5, 5.5 and 6.5 kpc.    

Water masers 1, 2, 4-7 and 9 have velocities which are comparable to the 
broadest feature of velocity profile observed by \citet{B06} and as such we 
associate these masers with the main cloud.  The velocity ranges of water 
masers 3 and 8 indicate that they are probably associated with a 
foreground cloud (or clouds) located at a distance of 1 kpc. Water maser 3 is separated by less than an arcsecond from the 22-GHz continuum source 
G\,333.060-0.489, indicating that 
the two are likely to be associated. This means that like water maser 3 this 
 continuum source is likely to be associated with a foreground cloud.  

Visual inspection of the \CO data of \citet{B06} shows that at velocities 
of approximately -10 \kms emission is restricted to a small range of right 
ascensions and declinations and is elongated almost perpendicular 
to the axis of the main GMC. Water maser 3 lies within this emission that is located at a 
distance of approximately 1 kpc. From the \CO data it appears unlikely 
that water maser 8 is associated with the same cloud as the \CO emission 
associated with this maser does not overlap with the \CO
emission associated with water maser 3 and the spatial separation between 
these masers is significant.

\subsubsection{Comments on individual sources}\label{sect:comments_sources}
1. {\em G\,332.653--0.621:} This water maser was discovered by
\citet{K76}, who observed it to have a peak flux density of 58 Jy at a
velocity of --47 \kms in 1975. Subsequent observations by \citet{B80}
during 1977 May showed a single maser feature at --44 \kms and a
slightly weaker intensity of 30 Jy.  We measured the peak flux density
to be 29 Jy at --45 \kmsns, similar to the observations made in
1977. No OH or 6.7-GHz methanol masers have been detected associated with this water maser \citep{Caswell,Simon96}.

This source is offset from the \ionhy region G\,332.663--0.621
identified by \citet{Huang99} by 35 arcseconds at a position angle of 58\degrees\/ and is situated
within the RCW 106 complex.  {\em IRAS} source {\em IRAS}\,16158--5055 is located 11 arcseconds away from this maser and 
 has colours typical
of an ultra-compact \ionhy region. This maser is separated from the
centre of the nearest \CO clump by 35 arcseconds and is 45
arcseconds from the centre of the 1.2-mm dust emission peak identified
by \citet{Mook04} at position angles of --79\degrees\/ and --83\degrees\/ respectively.

2. {\em G\,332.826--0.549:} This maser was discovered by \citet{BS82} in
1980 April, who reported it to have a peak flux density of 250 Jy at
--70.8 \kmsns. We found the intensity peak of the maser to be 239 Jy at
--59.1 \kms with emission covering a range of more than a 20 \kmsns. While the velocity range over which emission has been observed
has remained roughly constant since the maser's discovery, the
relative intensities and velocity of the strongest emission have not,
for example observations made by \citet{BS82} in 1981 May showed a peak
at --62 \kms of 198 Jy.

This maser appears to be associated with the {\em MSX} source
G\,332.8269--00.5489 which is offset by 5 arcseconds and may be also
associated with {\em IRAS}\,16164--5046 which is located 29 arcseconds
away, as well as a FIR source \citep{Karnik01}. The maser is probably
associated with a \CO clump identified by \citet{B06} and a 1.2-mm
dust emission peak identified by \citet{Mook04}, which are separated
from the maser by 11 arcseconds and 8 arcseconds respectively. This maser
is located about 2 arcminutes from the peak of the RCW 106B complex
which is centered on 16:20 --50:52 and is offset from the peak of the
22-GHz continuum source detected in our ATCA observations by 2
arcseconds. The water maser is separated from the OH maser G\,332.826-0.548 \citep{C98} by 8 arcseconds (see fourth sub-image of Fig.~\ref{fig:sick_pic}).

3. {\em G\,333.060--0.488:} This maser exhibits several spectral features
over an 18 \kms velocity range, with the most intense having a flux
density of 64~Jy at --8.7 \kmsns. This source was first observed at Mount
Pleasant on 8 August 2005 when the feature at --8.7 \kms was
approximately 6.5 Jy and the secondary feature at about 0 \kms was 3
Jy, implying a variation of a factor of 10 over an 11 month period. The velocity of this maser is comparable to one of the secondary features of the velocity profile of the GMC observed by \citep{B06} (see Fig.~\ref{fig:velocity_profile}). This maser is likely part of a different molecular cloud located along the line of sight at a distance of approximately 1 kpc (see Section~\ref{section:foreground_cloud}).

The ATCA observations detected a 22-GHz continuum source offset from
the water maser source by about an arcsecond (see third sub-image of Fig.~\ref{fig:sick_pic}). The water maser source
appears to be associated with GLIMPSE point source G\,333.0600--00.4888.

4. {\em G\,333.121--0.434:} This source was discovered by \citet{C74} in
1973 June. Subsequent observations made in 1976 August by \citet{B80}
showed a decline in the --51 \kms peak, while improved sensitivity
uncovered additional spectral features. We observed the current peak
intensity to be 161~Jy at --57.7 \kms with emission covering the
velocity range --60 to --46 \kmsns. The most prominent feature exhibited variation by a factor of two over a nine-month period.

This source is located within the RCW 106A structure and is offset from
two OH masers by about 50 arcseconds \citep{C98}. The 6.7-GHz methanol
maser G\,333.121--0.434 \citep{Simon05} is separated from the water
maser by less than an arcsecond and emission is seen over a similar
velocity range to the water maser. This maser falls within FIR, \CO
and 1.2-mm dust clumps with angular separations of 59, 40 and
47 arcseconds from the centre of the respective clumps. We detected
a 22-GHz continuum source separated from the water maser by
approximately 50 arcseconds which may be associated with the two OH
masers observed by \citet{C98} (see second sub-image of Fig.~\ref{fig:sick_pic}).

5. {\em G\,333.221--0.402:} The peak flux density of this maser has
remained approximately constant over the course of our observations,
however, initial observations made in 2005 August showed only one
spectral feature, while observations with the ATCA on 2006 July
detected four additional peaks. While three of these features can be
explained by the improved sensitivity offered by the ATCA the
remaining secondary peak of approximately 7 Jy should have been
detected in initial observations, suggesting this maser exhibits some
variability.

This water maser appears to be associated with GLIMPSE point source
G\,333.2205--00.4024 and is separated from the centre of the nearest
\CO clump by 33 arcseconds. This source falls within 10 arcseconds
of the centre of a 1.2-mm dust clump and 42 arcseconds from the
centre of a FIR source detected by \citet{Karnik01}.

6. {\em G\,333.29--0.38:} This water maser was discovered on 2005 June 26
when it was detected in both polarisations and in adjacent spectra,
however, when follow-up observations were made during 2005 September
and October the peak flux density of the maser was less than 1 Jy. The
earlier observations showed the maser to have a single spectral
feature of approximately 7 Jy at --49 \kmsns, comparable to the
velocity of the \ionhy region, GAL 333.3--00.4, \citep{Huang99} which
is situated 17 arcseconds away and has a velocity of --52.1
\kmsns. The {\em MSX} source G\,333.2898--00.3898 may be associated with the maser as the sources are separated by 33 arcseconds, less than the maser positional uncertainty of 1 arcminute. At the time of the observations made with the 
ATCA the peak flux density of this maser was less than the 5$\sigma$ detection limit of 0.75 Jy. 

7. {\em G\,333.364--0.358:} A decline in the peak flux density has been observed
since this maser was discovered on 2005 August 25. In the initial
observations a peak flux density of 9 Jy was observed compared to the
final observations made with the ATCA where a peak flux of 3.2 Jy was
recorded. This source is separated from the centre of the nearest \CO
clump by 47 arcseconds and the centre of a FIR source detected by
\citet{Karnik01} by 37 arcseconds.

8. {\em G\,333.428--0.380:} This maser consists of two main spectral
features at velocities of --4.5 and 4.1 \kms with flux densities of 4
Jy and 12.4 Jy respectively. A decrease in the flux density of the
primary feature has been observed since observations made during 2005
April when it had a peak flux density of 26 Jy. This is the only maser
detected in our survey for which there is no {\em IRAS}, {\em MSX},
1.2-mm dust or other maser species within 2 arcminutes. There is however
a nearby \CO emission peak located at about 40 arseconds from the
maser. Like water maser G\,333.060--0.488, the velocity of this maser indicates that it probably belongs to another molecular cloud located at a distance of approximately 1 kpc \citep{B06} (see Section~\ref{section:foreground_cloud}). 

9. {\em G\,333.608--0.215:} This is one of the earliest water
masers to be discovered and was first detected by \citet{J72} who observed a
primary feature at --49 \kms and a secondary feature at --57 \kmsns.
Observations in 1976 August by \citet{B80} found a 100 Jy peak at --52
\kmsns. We detected emission over a 26 \kms velocity range with the most
intense feature of 103 Jy at --49.2 \kms and a decline in the --52 \kms
feature to about 38 Jy. There is an associated OH maser observed by
\citet{C98} which is displaced from the water maser by less than an arcsecond
and shows emission over a velocity range of --48 \kms to --36\kms (see first sub-image of Fig.~\ref{fig:sick_pic}).

This maser is associated with the well known \ionhy region
G\,333.6--0.2 which is almost totally obscured at optical wavelengths
but is one of the brightest objects at longer wavelengths. This maser
is probably associated with an {\em IRAS} source, FIR source, \CO
clump, CS emission and 1.2-mm dust clump.

\begin{table*}
	\caption{Summary of all possible associations. Details of GLIMPSE, {\em IRAS}, {\em MSX}, FIR, 1.2-mm dust, \CO and CS sources can be seen in Tables~\ref{tab:glimpse}, ~\ref{tab:sep}, ~\ref{tab:FIR}, ~\ref{tab:dust}, ~\ref{tab:CO} and ~\ref{tab:CS}. y indicates the presence of an association, n indicates that there is no association, * indicates the \CO emission peaks that we identify that were outside the velocity range of clump analysis carried out by \citet{B06} and - indicates that no data over a similar velocity to the respective water masers was available.}
	\begin{tabular}{ccccccccccc}\hline
	{\bf Source} & {\bf GLIMPSE} & {\bf {\em MSX}} & {\bf {\em IRAS}} & {\bf FIR} &{\bf 1.2 mm}& {\bf \CO} & {\bf CS} & {\bf Methanol} &{\bf OH}\\
	{\bf number} & & & &&{\bf dust} & & & {\bf maser} & {\bf maser}\\ \hline
	1 & n & n & y & n & y& y & y & n & n\\
	2 & y & y & y & y &y &  y & y & n & n\\
	3 & n& n & n & n &y &  y* & - & n & n\\
	4 & y & n & n & y & y& y & y & y & n\\
	5 & n & n & n & y & y& y & y &n & n\\
	6 & y & y & n & y & y& y & y & n & n\\
	7 & y & n  & n & n &n & y & y & n&n \\
	8 & n & n & n & n & n& y* & - & n &n \\
	9 &n & n & y &y & y& y & y & n & y\\ 
	\hline
	\end{tabular}
\label{tab:ass}
\end{table*}

\subsection{22-GHz continuum sources}\label{section:cont}

Four radio continuum sources were detected in the ATCA observations.
Their properties are summarised in Table~\ref{tab:continuum} and are
shown in the sub-images of Fig.~\ref{fig:sick_pic}. As there is no 
high resolution, low frequency continuum data available for the region,
spectral indices (and therefore likely source mechanisms) could not be 
determined.

\begin{table*}
  \caption{22-GHz continuum sources. Column 1 is the source number of the 
    nearest water maser, Columns 2-4 give the position of the continuum 
    source, column 5 gives the peak of the continuum source in mJy/beam, 
    column 6 gives the total flux density of the continuum source in mJy and 
    column 7 gives the angular separation between the continuum source and the
    nearest water maser source.}
  \begin{tabular}{ccccccc}\hline
	{\bf Source} & {\bf Continuum} & {\bf Right Ascension} &{\bf Declination} & {\bf {\em F}$_{Peak}$} & {\bf Total} & {\bf Separation}\\
	{\bf number} & {\bf Source} & {\bf (J2000)}&{\bf (J2000)}& {\bf (mJy/beam)} &{\bf Flux Density} &  {\bf from maser}\\
	& {\bf ($l,b$)}& & & & {\bf (mJy)} & {\bf (arcsec)}\\ \hline
	2& G\,332.826--0.549&16:20:11.089 & --50:53:14.07& 948 &1180 & 2\\
	3& G\,333.060--0.489& 16:20:58.002&--50:40:47.32& 91 & 126 &  1\\
	4& G\,333.135--0.432& 16:21:03.017&--50:35:12.54& 720 & 901 & 50\\
	9 & G\,333.605--0.212 &16:22:09.605&--50:05:59.98& 569 & 631  &15\\ \hline
	\label{tab:continuum}
	\end{tabular}
\end{table*}

{\em G\,333.826--0.549} This is the strongest radio continuum source
detected in our ATCA observations and its peak is located within
2~arcseconds of the water maser G\,332.826--0.549. The nearest infrared
source is {\em IRAS}\,16164--5046, with which it is possibly associated.

{\em G\,333.060--0.489} This is the weakest continuum source that we
detected. The source is spatially coincident with GLIMPSE point source
G\,333.0600--00.4888 as well as the water maser G\,333.060--0.488. As the
separation between the continuum source and the near-by water maser is less than 
an arcsecond it seems likely that the two are associated. As this water maser (source number 3) 
is probably associated with a foreground cloud (see Section~\ref{section:foreground_cloud}) and not the 
main GMC, it is likely that this continuum source also belongs to the foreground cloud.

{\em G\,333.135--0.432} This continuum source, while not associated with
any water masers, does appear to be associated with two OH masers
observed by \citet{C98} (shown in Fig.~\ref{fig:sick_pic}). There is
no associated infrared source.
 
{\em G\,333.605--0.212} This continuum source is coincident with the
{\em MSX} source G\,333.6046--00.2124 and {\em IRAS} source {\em IRAS}\,16183--4958. These infrared sources are separated from the the peak of the continuum
emission by 1.3 and 5.8 arcseconds respectively. This source is offset from
the nearest water maser that we observe by about 15 arcseconds.

\section{Discussion} \label{sec:discussion}

\subsection{Association with other maser species}

 To our knowledge, this is
the first high-mass star formation complex for which untargeted
searches have been made in all of main-line OH, 6.7-GHz methanol and
22-GHz water masers.  Previous targeted searches for water masers
towards known 6.7-GHz methanol masers achieved detection rates of
around 50 percent \citep*{Szymczak2005}. In contrast we find that only
25 percent of the 6.7-GHz methanol masers that fall within our survey
region have an an associated water maser. We additionally find that
only 11 percent of the water masers that we detect have an associated
methanol maser which implies that targeted water maser searches
towards 6.7-GHz methanol masers may not be the most efficient way to
increase the number of known water masers.  It also supports the
hypothesis that water masers may be the most prevelant species within
star formation regions as our relatively insensitive survey has
detected twice as many water masers as either 6.7-GHz methanol or OH
in the corresponding regions.

\subsection{Association with infrared sources}

\citet{Ellingsen2006} found that approximately two-thirds of 6.7-GHz
methanol masers have an associated GLIMPSE point source, and less than
10 percent of sources are not associated with mid-infrared emission
(at the sensitivity of the GLIMPSE observations).  A search of the
GLIMPSE catalogue reveals four of the water maser sources detected
(all of them new discoveries) have an associated GLIMPSE point source
within 2 arcseconds. The details of these GLIMPSE point sources are
summarized in Table~\ref{tab:glimpse}.  Of the remaining five water
masers, the four previously detected sources are all clearly projected
against regions of bright mid-infrared emission (see
Fig.~\ref{fig:3colour}).  \citet{Ellingsen2006} suggested that a
search for 6.7-GHz methanol masers towards GLIMPSE point sources
meeting the criteria [3.6]--[4.5] $>$ 1.3 mag and 8.0~\micron\/
magnitude $<$ 10 would detect more than 80 percent of this class of
maser.  From Table~\ref{tab:glimpse} it can be seen that three of the
four GLIMPSE point sources associated with water masers satisfy these
criteria.  We compared the [3.6]--[4.5] colours of the GLIMPSE sources
associated with water masers with the colours of GLIMPSE sources associated with 6.7-GHz methanol
masers (see fig.\ 16 of \citeauthor{Ellingsen2006}).  The colours of the sources associated with water masers
are clustered in the less red end of the range observed in the methanol
associated sources.  The idea that different maser species may trace
different phases of the high-mass evolutionary sequence is not new,
however, it has been receiving renewed attention lately, as
sensitive, high-resolution observations at sub-millimetre through
mid-infrared wavelengths become more readily available.
\citet{Ellingsen2006} looked at this question in some detail and we
will not repeat the arguments here. However, the striking difference
between the relative location of the water and methanol masers in the
G\,333.2--0.6 GMC, and the less red colours of the water maser
associated GLIMPSE sources supports the hypothesis that 6.7-GHz
methanol masers may trace a generally earlier evolutionary phase than
water masers.  The very small water maser sample size prevents us from
drawing any firm conclusions, however, it suggests that comparison of
the properties of GLIMPSE sources associated with water and methanol
masers may provide useful insights relating to this question.

The top and bottom inset images in Fig.~\ref{fig:sick_pic} each show a
ring of diffuse 8.0-$\mu$m emission surrounding a darker region.  We
have closely examined the GLIMPSE images of these regions. In the case of the southern most region
(associated with the water maser G\,333.826--0.549) it is clear that this is the result of a known image artifact that
is present in some GLIMPSE sources where saturation occurs. 
An explanation for the northern most source (associated with the water maser G\,333.608--0.215) is not 
as clear. The ring like structure  seen in the 8.0-$\mu$m is similarly seen in the other GLIMPSE bands with radius decreasing with wavelength and the point where the dip in flux density per steradian occurs is close to the GLIMPSE saturation level. Interestingly, the {\em MSX} images of the region show a similar structure, warranting further observations of this source at both radio and infrared wavelengths to better understand its nature.

\begin{table*}
  \caption{Water masers with GLIMPSE point source associations, here IRAC bands 1, 2, 3 and 4 correspond to 3.6-$\mu$m, 4.5-$\mu$m, 5.8-$\mu$m and 8.0-$\mu$m. Column 1 gives the water maser source number (see Table~\ref{tab:masers} for details), column 2 is the associated GLIMPSE point source, column 3 gives the angular separation between the GLIMPSE point source and the water maser source and colums 4 to 7 show the magnitudes of IRAC bands 1, 2, 3 and 4 for each of the GLIMPSE point sources, while column 8 gives the [3.6]--[4.5] colour for each.}
\begin{tabular}{cccccccc}\hline
	{\bf Source} & {\bf GLIMPSE} & {\bf Separation} & & {\bf Magnitude} & {\bf (mag)} & & {\bf [3.6]--[4.5]} \\
	{\bf number}& {\bf point source} & {\bf (arcsec)} & {\bf IRAC} & {\bf IRAC} & {\bf IRAC} & {\bf IRAC} & {\bf colour}\\
	& {\bf ($l,b$)} & & {\bf  band 1} & {\bf band 2} & {\bf  band 3} & {\bf  band 4} & {\bf (mag)}\\ \hline
	3 & G\,333.0600--00.4888 & 1.3 & 11.964 & 10.633 & 9.258 & - & 1.331\\
	5 & G\,333.2205--00.4024 &1.8  & 8.973 & 7.434 & 6.368 & 5.895 & 1.539\\
	6 & G\,333.3639--00.3574 & 0.8 & 11.263 & 9.780 &8.560 & 7.551& 1.483\\
	7 & G\,333.4285--00.3809 & 2.0& 14.233 & 13.373 & - & - &0.860\\ \hline
	\end{tabular}
	\label{tab:glimpse}
\end{table*}

Three of the nine water masers detected in this survey (all previously
known sources) have an {\em IRAS} source within 30 arcseconds. Two of the water masers that we observe have an associated {\em MSX} source. The details of these {\em IRAS} and {\em MSX} sources
are presented in Table~\ref{tab:sep}.

\begin{table*}
  \caption{Possible infrared and water maser source associations. Column 1 gives the water maser source number (see Table~\ref{tab:masers} for details), column 2 shows the closest {\em IRAS} source within 2 arcminutes of the water masers, column 3 gives the angular separation between the water maser sources and the {\em IRAS} source, column 4 is the closest {\em MSX} source to within 1 arcminute and column 5 gives the angular separation between the {\em MSX} source and water maser source.}
\begin{tabular}{ccccccc}\hline
	{\bf Source} & {\bf {\em IRAS}} & {\bf Separation} & {\bf {\em MSX}} & {\bf Separation} \\
	{\bf number} & {\bf source} & {\bf (arcsec)} & {\bf source} & {\bf (arcsec)} \\ \hline
	1 & 16158--5055 & 11 &\\ 
	2 & 16164--5046 & 29 & G332.8269--00.5489 & 5 & \\
         6& - & & G333.2898--00.3898& 33&  \\
         9& 16183--4958& 20 \\ \hline 	
         \label{tab:sep}
	\end{tabular}
\end{table*}

\begin{table*}
\caption{Separation from FIR sources, detection at 150 and 210 $\mu$m. Positions given are of the peak of the 210 $\mu$m \citep{Karnik01}. Columns are as follows; 1 is the maser source number, 2 is the FIR source number quoted by \citet{Karnik01}, 3 and 4 are the Right Ascension and Declination of the clumps, 5 gives the angular separation between the water maser and the peak of the FIR source, columns 6 and 7 give the FIR source's peak flux densities at 150 and 210 $\mu$m respectively and column 8 gives the source luminosity.}
\begin{tabular}{cccccccc}\hline
	{\bf Source} & {\bf FIR} & {\bf Right}&{\bf Declination} & {\bf Separation} & {\bf 150 $\mu$m} & {\bf 210 $\mu$m} & {\bf Luminosity}\\
	{\bf number} & {\bf source} & {\bf Ascension} & {\bf (J2000)}  & {\bf (arcsec)} & {\bf Flux Density} & {\bf Flux Density} & {\bf (10$^{3}$ L$_{\odot}$)}\\
	& {\bf number} &  {\bf (J2000)} &  & & {\bf (Jy)} & {\bf (Jy)}\\ \hline
	2 & S11 & 16:20:06.0 & --50:53:19 & 48& 15380 & 7790 & 411\\
	4& S15 & 16:20:57.9 & --50:34:55 & 59& 20811 & 10009 & 482\\
	5& S18 &16:21:20.3 & --50:29:43 & 42 & 2180 & 1043 & 115\\
	6 & S20 &16:21:29.8 & --50:25:57 & 37 & 15572 & 8015 & 460\\
	9 & S23 & 16:22:08.9 & --50:05:28 & 35 & 27841 & 11555 & 921\\ \hline
         \label{tab:FIR}
	\end{tabular}
\end{table*}

\citet{Karnik01} detected 23 FIR sources in a survey of the GMC at 150
and 210 $\mu$m. Of the six most luminous of these sources, five have
associated water masers (see Table~\ref{tab:FIR} for details).

\subsection{Association with molecular gas and dust clumps}

\subsubsection{\CO clump associations}

\begin{figure}
  \epsfig{file=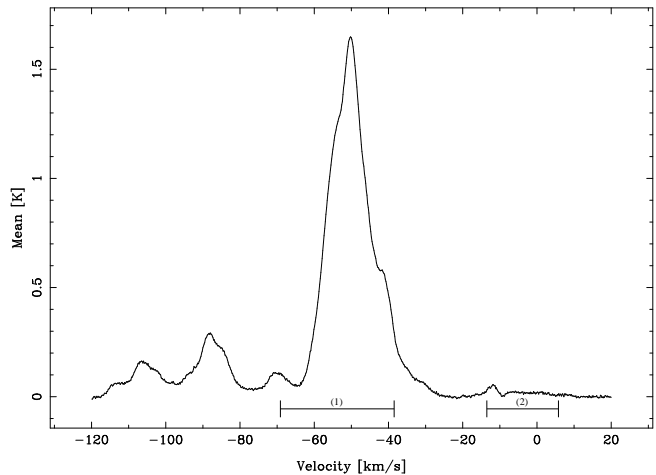,height=0.35\textwidth}
  \caption{The mean velocity profile of the \CO emission averaged over
    the entire field of observations from \citet{B06}. Overlaid are
    the velocity ranges of the water masers. (1) indicates the total
    velocity range of the seven water masers with comparable velocity
    ranges; maser sources 1, 2, 4, 5, 6, 7 and 9, while (2) indicates
    the velocity ranges of maser sources 3 and 8.}
  \label{fig:velocity_profile}
\end{figure}

\citet{B06} produced {\scshape{clumpfind}} fits of the GMC using the 2D \CO
data integrated over the velocity range --65 \kms to --35 \kms which
covers the most prominent feature of the velocity profile seen in
Fig.~\ref{fig:velocity_profile}. Seven of the nine water masers that we detect fall within the velocity range analysed by \citet{B06} and all of these masers fall within a \CO clump that they identify.

We obtained the \CO data of \citet{B06} in the velocity range --120
\kms to 20 \kms which has allowed us to identify additional \CO
emission peaks (or possible clumps) which may be associated with the
two masers not coincident with a clump identified by \citet{B06},
G\,333.060--0.488 and G\,333.428--0.380. These emission peaks, like
their associated water masers, have velocity ranges which fall outside
that analysed by \citet{B06}. Water masers G\,333.060--0.488 and
G\,333.428--0.380 show emission over a comparable velocity range of a
secondary feature seen in the \CO emission
(Fig.~\ref{fig:velocity_profile}).

Details of the \CO clumps identified by \citet{B06} and the \CO
emission peaks that we identify along with their associated water
maser sources are summarized in Table~\ref{tab:CO}, while clump
properties identified by \citet{B06} are presented in
Table~\ref{tab:CO2}. The velocity ranges of the \CO clumps or emission
peaks have been obtained from visual inspection of the data. The \CO 
emission peaks have
more extensive velocity ranges than the associated water maser emission
in all cases with the exception of water masers G\,333.060--0.488 and
G\,333.428--0.380. Interestingly these two water masers have similar
velocity ranges which place them both at a distance of around 1 kpc (see Section~\ref{section:foreground_cloud}). In the
majority of cases the velocity of the peak flux density of the water
masers are within a few \kms of the velocity of the associated peak
\CO emission.

\begin{table*}
 \caption{Water maser sources with nearby \CO clumps identified by \citet{B06} (over velocity range --65 \kms to --35 \kms) or emission peaks as identified by visual inspection of the data used by \citet{B06}. Column 1 is the water maser source number (see Table~\ref{tab:masers} for details), column 2 and 3 give the velocity range and the velocity of the peak emission of the water maser, column 4 gives the \CO clump number from \citet{B06} if applicable, columns 5 and 6 show the right ascension and declination of the \CO clump or alternatively the \CO emission we identified from the \CO data from \citet{B06}, columns 7 and 8 give the velocity range and the velocity of the \CO emission peak as we have identified and column 9 is the separation between the centre of the clump that \citet{B06} reported or the emission peak that we identified. The velocity ranges of the additional emission peaks that we identify as being associated with water masers 3 and 8, indicate that they are probably located along the line of sight at a distance of approximately 1 kpc \citep{B06}.}

\begin{tabular}{ccccccccc}\hline
	{\bf Source}& {\bf Velocity} & {\bf Peak} & {\bf \CO} & {\bf Right} & {\bf Declination} & {\bf Velocity} & {\bf Peak}&{\bf Separation}\\
	{\bf number} & {\bf Range} & {\bf Velocity} &{\bf source} & {\bf Ascension} & {\bf (J2000)} & {\bf Range} & {\bf Velocity} & {\bf (arcsec)}\\ 
	& {\bf  (km s$^{-1}$)} & {\bf (\kmsns)}& {\bf number} & {\bf (J2000)} & &  {\bf (kms$^{-1}$)} &  {\bf (kms$^{-1}$)}\\ \hline
	1 & --59,--43& --45.3 &9 & 16:19:40 & --51:03:27 &--60,--43 & --50 & 35\\
	2 & --69,--45 & --59.1 & 7 & 16:20:11 & --50:53:27&--65,--44 & --55  & 11\\
         3 & --13,5 & --8.7&- & 16:20:58 & --50:40:39 &--9,--2 & --6 & 8\\
         4 & --60,--46& --57.7&1 & 16:21:03 & --50:35:27 & --64,--45 &--51& 40 \\
         5 &--58,--48& -52.0 &14 & 16:21:21 & --50:30:03 & --57,--40&--52 & 33\\
         6  & --58,--47  & --49.0 &  2 & 16:21:31 & --50:26:51 & --58,--45& --52& 23\\
         7 &--55,--49& -52.6 &22 & 16:21:42 & --50:21:39 & --67,--42& --51& 48\\
         8& --6,6& 4.1&- & 16:22:05 & --50:21:02 & --1,10 & 6& 40\\
         9 & --64,--38& --49.2&5 & 16:22:08 & --50:06:27& --61,--32 & --46& 42\\ \hline
         \label{tab:CO}
	\end{tabular}
\end{table*}

\begin{table*}
\caption{ Properties of the \CO clumps \citep{B06}. Column 1 is the number of the associated water maser source, column 2 is the \CO clump number \citep{B06}, column 3 is the integated flux density of the clump peak, column 4 is the \CO clump peak in terms of antenna temperature, column 5 is the clump radius, column 6 is the \CO column density and column 7 is the total LTE molecular mass.}
	\begin{tabular}{cccccccc}\hline
	{\bf Source} & {\bf \CO}& {\bf Integrated Flux} & {\bf Peak} & {\bf Radius} & {\bf N(\COns)} & {\bf Mass}\\
	{\bf number} & {\bf Clump} & {\bf (10 K \kmsns)} & {\bf (K)} & {\bf (pc)} & {\bf (10$^{16}$ cm$^{-2}$)} & {\bf (10$^{3}$ M$_{\odot}$)}\\ \hline
	1&9 &10.5 & 13.3 & 1.7 & 6.6 & 4.6\\
	2 & 7 & 12.0 & 14.4 & 1.8 & 6.2 & 5.0\\
	4 & 1 & 18.5 & 46.0 & 2.6 & 10.1 & 16.0\\
	5 & 14 &8.8 &15.7 & 1.8 & 6.9 & 5.5\\
	6 & 2 & 15.8 & 41.4 & 2.4 & 10.3 & 14.4\\
	7 & 22 & 8.9 & 25.0 & 2.2 & 7.6 & 8.7\\
	9 & 5  & 13.8 & 27.4 & 2.2 & 7.9 & 9.5\\ \hline
	\label{tab:CO2}
	\end{tabular}
\end{table*}

\subsubsection{Associations with CS emission}

\begin{table*}
 \caption{Water masers sources with nearby CS emission peaks. Column 1 is the water maser source number, columns 2 and 3 give the velocity range and velocity of emission peak, columns 4 and 5 are the right ascension and declination of the CS emission peak, columns 6 and 7 show the velocity range and the emission peak of the CS emission and column 8 shows the angular separation between the water maser source and the CS emission peak.}

\begin{tabular}{ccccccccccc}\hline
	{\bf Source}& {\bf Velocity} & {\bf Velocity} & {\bf CS} & {\bf Clump} & {\bf Velocity} & {\bf Peak}& {\bf Separation}\\
	{\bf number} & {\bf Range} &  {\bf Peak} & {\bf Right Ascension} & {\bf Declination} & {\bf Range} & {\bf Velocity} & {\bf (arcsec)}\\ 
	& {\bf (kms$^{-1}$)} & {\bf (kms$^{-1}$)}& {\bf (J2000)} & {\bf (J2000)} &  {\bf (kms$^{-1}$)} & {\bf (kms$^{-1}$)} \\\hline
	1 & --59,--43& --45.3 & 16:19:40 & --51:03:30&--56,--44 & --50&33 \\
	2 & --69,--45 & --59.1 &16:20:11 & --50:53:19&--61,--51&--55 & 4\\
         4 & --60,--46 & --57.7 &16:21:03 & --50:34:56& --61,--44& --53& 63\\
          5&--58,--48& --52.0 & 16:21:22 & --50:30:46&--57,--48 & --51& 45\\
         6  & --58,--47  & --49.0 & 16:21:29 & --50:26:31& --58,--45 & --51& 12\\
         7 &--55,--49 & --52.6 & 16:21:41 & --50:23:19 &--55,--45 & --49& 69\\
         9 & --64,--38&--49.2 & 16:22:08 & --50:06:17& --58,--39 & --47& 40\\ \hline	
         \label{tab:CS}
	\end{tabular}
\end{table*}

The GMC has been  mapped in CS emission by the Delta Quadrant Survey Team as part of the ongoing project
at University of New South Wales. CS emission is sensitive to higher densities and as a result the mapped region is less extensive than that covered in the \CO mapping (and has been
obtained in advance of publication as a result of correspondence with
Nadia Lo from UNSW).  This means that CS data was not available for two
of the positions of the detected water masers. We inspected the data
over a velocity range of --80 \kms to --20 \kms and identified CS
emission peaks near all of the seven water masers for which CS data
was available. The details of these emission peaks and their
associated water masers are summarized in Table~\ref{tab:CS}.

Unlike the \CO emission, CS emission is observed over a smaller
velocity range than the associated water maser emission, however like
the \CO emission the velocity of the peak flux density of the water
masers correlates to within a few \kms with the velocity of the CS
emission peaks. The association of the water masers with CS emission
indicates that the water masers are probably associated with massive
star formation \citep{BNM96}.

\subsubsection{Association with 1.2-mm dust clumps}

Seven of the water masers that we detect are associated with a 1.2-mm
dust emission peak observed by \citet{Mook04} and the properties of
the dust clumps are summarized in Table~\ref{tab:dust}. \citet{Mook04} identified 95 1.2-mm dust clumps within the GMC and 73 of these fall within our observed regions. A statistical analysis of the 1.2-mm clumps that fall within the observed regions is presented in Section~\ref{section:stats}.

\begin{table*}
  \caption{Maser sources with nearby 1.2-mm dust clumps identified by \citet{Mook04}. Column 1 is the water maser source number, column 2 is the dust clump number, columns 3 and 4 give the right ascension and declination of the dust clump, column 5 is the angular separation between the water maser and associated dust clump, column 6 gives the peak flux density for the clump, column 7 gives the clump radii, column 8 is the total integrated flux densities of the clumps, column 9 gives the estimated mass of the clumps assuming a temperature of 40 K and column 10 gives the column density.}

\begin{tabular}{cccccccccc}\hline
	{\bf Source} & {\bf 1.2-mm} & {\bf Right} & {\bf Declination} &  {\bf Separation} & {\bf F$_{peak}$} & {\bf Radius} & {\bf F$_{v}$} & {\bf Mass} & {\bf n$_{H_{2}}$}\\
	{\bf number} & {\bf dust} & {\bf Ascension} & {\bf (J2000)} & {\bf (arcsec)} & {\bf (mJy/beam)} & {\bf (pc)} & {\bf (Jy)} & {\bf (M$_{\odot}$)} & {\bf (10$^{4}$ cm$^{-3}$)} \\ 
	& {\bf source}& {\bf (J2000)}\\ \hline
	1 & MMS84 & 16:19:38.9 & --51:03:28 & 45 & 2499& 1.48 & 21.9 & 2871 & 2.45\\
	2 & MMS68 & 16:20:11.9 & --50:53:17 & 8 & 12460& 1.70 & 36.1& 5548 & 2.56\\
         3 & MMS51 & 16:20:52.1 & --50:40:51 & 56 &348 & 1.11& 2.7 & 1029 & 0.73\\
         4 & MMS39 & 16:21:03.7 & --50:35:23 &47& 8925 & 1.19 & 32.6 & 4584 & 6.18\\
         5 & MMS33&16:21:18.6 & --50:30:25 & 10 & 1095& 0.90 & 3.5 & 1325& 1.76\\
         6 & MMS29 & 16:21:32.7 & --50:27:12 & 44 & 6502 & 1.27 & 24.7 & 9281 & 4.38\\
        9 & MMS5& 16:22:10.1 & --50:06:06 & 14 & 40013& 1.88 & 129.5 & 15936 & 5.44\\ \hline	
         \label{tab:dust}
	\end{tabular}
\end{table*}

\subsection{\CO and 1.2-mm dust clump analysis}\label{section:stats}

In order to investigate the properties of the molecular gas and dust
in the regions with associated water maser emission we have fitted a
Binomial generalized linear model (GLM) \citep{Mccul} to the maser
presence/absence data using \CO and 1.2-mm dust clump properties
reported by \citet{B06} and \citet{Mook04} respectively, as
predictors. In the case of the \CO clumps the properties considered
were the integrated flux densities of the clump peaks (10 K \kmsns), the peak flux in terms of antenna temperature (K), clump radius (pc), \CO column
density (10$^{16}$ cm$^{-2}$) and the total LTE molecular mass
calculated from the \CO data. In the case of the 1.2-mm dust clump
analysis the properties F$_{peak}$ (mJy/beam), radius (pc), total integrated
flux density (F$_{v}$) in Jy, mass (M$_{\odot}$) and {\em n}$_{H_{2}}$ (10$^{4}$ cm$^{-3}$)
where investigated as predictors. A Binomial GLM predicts the
probability, {\em p}$_{i}$, of finding a maser in the {\em i}$^{th}$
clump, in terms of the clump properties {\em x}$_{1i}$ {\em x}$_{2i}$
{\em x}$_{3i} \ldots$ {\em x}$_{mi}$. The model takes the form

\begin{eqnarray*}
  y_{i} & \sim & \mbox{Bin}(1,p_{i})\\
  \mbox{log}{\frac{p_{i}}{1-p_{i}}} & = &\beta_{0} + \beta_{1}x_{1i} + \beta_{2}x_{2i} + \ldots + \beta_{m}x_{mi}
\end{eqnarray*}
where $y_{i}$ is the maser presence or absence in the {\em i}$^{th}$
clump and $\beta_{0}$, $\beta_{1}$, $\beta_{2} \ldots \beta_{m}$
are the regression coefficients to be estimated.

To test the significance of individual clump properties all possible
single term models were fitted, and compared by analysis of deviance
to the null model consisting of only an intercept. Stepwise model
selection based on the Akaike Information Criteria (AIC) \citep{Burnham}
was used to select the most parsimonious model with the greatest
predictive power. The AIC is a trade off between goodness of fit and
model complexity and is defined as

\begin{eqnarray*}
AIC=-2(\mbox{max log likelihood})+2(\mbox{number of parameters})
\end{eqnarray*} 
with the preferred model being the model with the lowest AIC
\citep{MASS}.

For ease of comparison between the data sets box plots were created for each of the clump properties. The solid horizontal line in each of these plots represents the median of the data. The box represents the 25$^{th}$ to the 75$^{th}$ percentile, while the vertical line from the top of the box represents data from the 75$^{th}$ percentile to the maximum value and the vertical line from the bottom of the box represents data from the 25$^{th}$ percentile to the minimum value. Outliers are represented by dots. Box plots of each of the clump properties of both the \CO clumps and the 1.2-mm dust clumps are shown in Figs.~\ref{fig:CO_box} and \ref{fig:dust_box} respectively.

\subsubsection{\CO clump results}

\begin{figure}
\includegraphics[angle=270,scale=0.7]{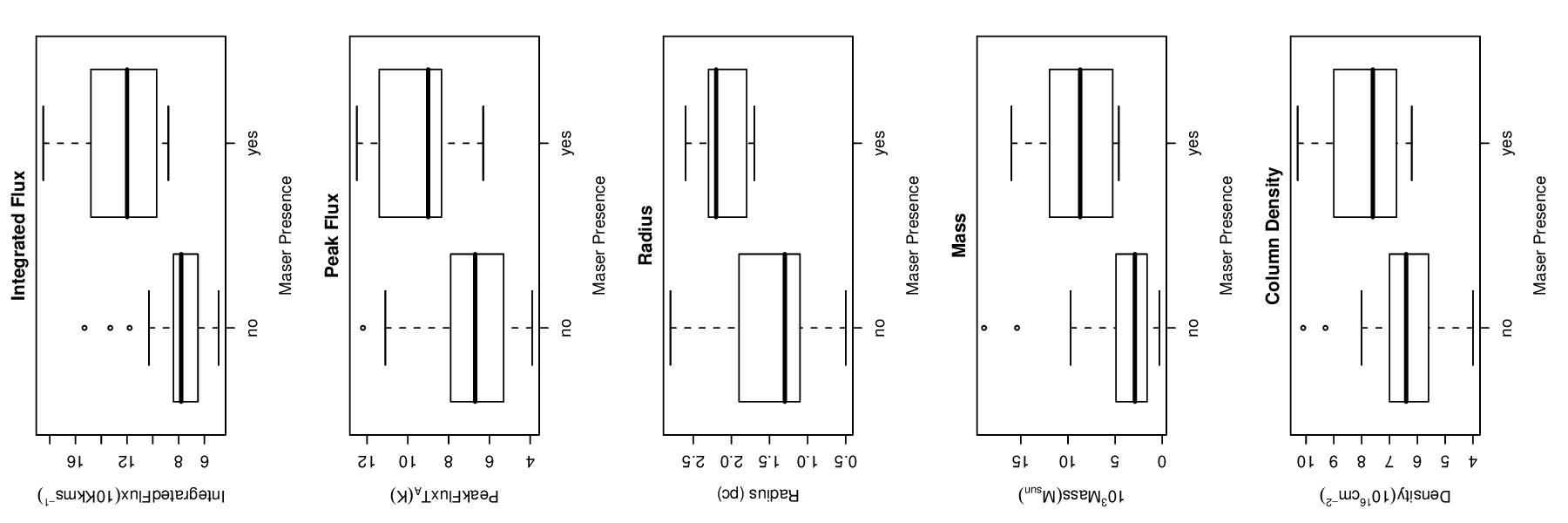}
\caption{Box plots of each of the \CO clump properties split into the categories of yes and no, referring to maser presence and absence respectively.}
\label{fig:CO_box}
\end{figure}

Fits of the single term addition Binomial model to the \CO clump
properties reported by \citet{B06}, showed an increasing probability of the presence of a
maser was associated with all of the tested factors (the integrated flux density of the clump peak, \CO clump peak in terms of antenna temperature,
clump radius, \CO column density and the total LTE molecular mass
calculated from the \CO data). This means that any of the clump
properties (in isolation) gives an indication of the likelihood of
maser presence. Table~\ref{tab:co_sta} gives a summary of the single
term addition Binomial model. The same information is shown
graphically in Fig.~\ref{fig:CO_box} which clearly illustrates that
for all clump properties there is a difference between the \CO clumps
that have an associated water maser and those that do not. In general
the \CO clumps with associated water masers are bigger, denser,
brighter and more massive. The most parsimonious model for predicting maser presence
involved the integrated flux density of the clump peak, clump radius and the total LTE
molecular mass calculated from the \CO data. This means that
if the integrated flux density of the peak, radius and mass is known for a \CO clump
then a probability of maser presence can be determined. The estimated regression relation is 
\begin{eqnarray*}
\mbox{log}{\frac{p_{i}}{1-p_{i}}}=-21.2018+1.3037x_{integrated}+8.0589x_{radius}\\-1.2290x_{mass},
\end{eqnarray*}
where {\em x}$_{integrated}$, {\em x}$_{radius}$ and {\em x}$_{mass}$ represent clump properties integrated flux density of the clump peak (10 K \kmsns), radius (pc) and mass (10$^{3}$ M$_{\odot}$). The full regression summary is shown in Table~\ref{tab:CO_regression}.

\begin{table}
  \caption{Analysis of deviance table for the single term models (using \CO clump properties), showing the AIC and the deviance together with the associated likelihood ratio statistic and p-value for the test of the hypothesis that the stated single term model provides no better fit than the null model consisting only of an intercept.}
\begin{tabular}{ccccc}\hline
	{\bf Predictor} & {\bf AIC} & {\bf Deviance} & {\bf LRT} & {\bf p-value}\\ \hline
	{\bf none} &39.098 & 37.098\\
	{\bf Integrated} &28.774  &24.774 &12.325& 0.000447 \\
	{\bf Peak} &33.093&    29.093  & 8.005 &0.004664\\
         {\bf Radius} &34.063  & 30.063   &7.035 &0.007994\\
         {\bf Density}   &34.615   &  30.615  & 6.483 &0.010892\\
         {\bf Mass}&35.265&      31.265  & 5.833 &0.015729\\ \hline
         \label{tab:co_sta}
	\end{tabular}
\end{table}

\begin{table}
	\caption{Summary table for the Binomial regression model, showing for each predictor the estimated coefficient and its standard error, and the standardised z-value and p-value for the test of the hypothesis that $\beta_{i}$=0.}
\begin{tabular}{ccccc}\hline
	{\bf Predictor}& {\bf Estimate} & {\bf Std. Error} & {\bf z value} & {\bf p-value}\\ \hline
	{\bf Intercept} & --21.2018 & 9.0538 & --2.342 & 0.0192\\
	{\bf Integrated} & 1.3037 & 0.5041 & 2.586 & 0.0097\\
	{\bf Radius} & 8.0589 & 5.0069 & 1.610 & 0.1075\\
	{\bf Mass} & --1.2290 & 0.6153 &  --1.997 & 0.0458\\ \hline
         \label{tab:CO_regression}
	\end{tabular}
\end{table}

Setting the threshold probability of the \CO clump model at 0.5
(i.e. a value greater than 0.5 suggests a water maser will be
associated with a clump, while a lower value suggests no water maser),
we find the misclassification rates to be low. Of the 40 \CO clumps
that fall in our observation regions which correspond the the main part of the GMC ($-65$ \kmsns $\leq$ v$_{\rm LSR}$ $\leq$ $-35$ \kmsns), seven have an associated water
maser while our model predicts that five of these have an associated
water maser and returns a false-negative for the remaining two
clumps. There are 33 clumps within the survey regions that do not have
associated water masers, our model predicts that 31 of these do not
have associated water masers but returns a false-positive for the
remaining two clumps.

\subsubsection{1.2-mm dust clump results}

\begin{figure}
\epsfig{file=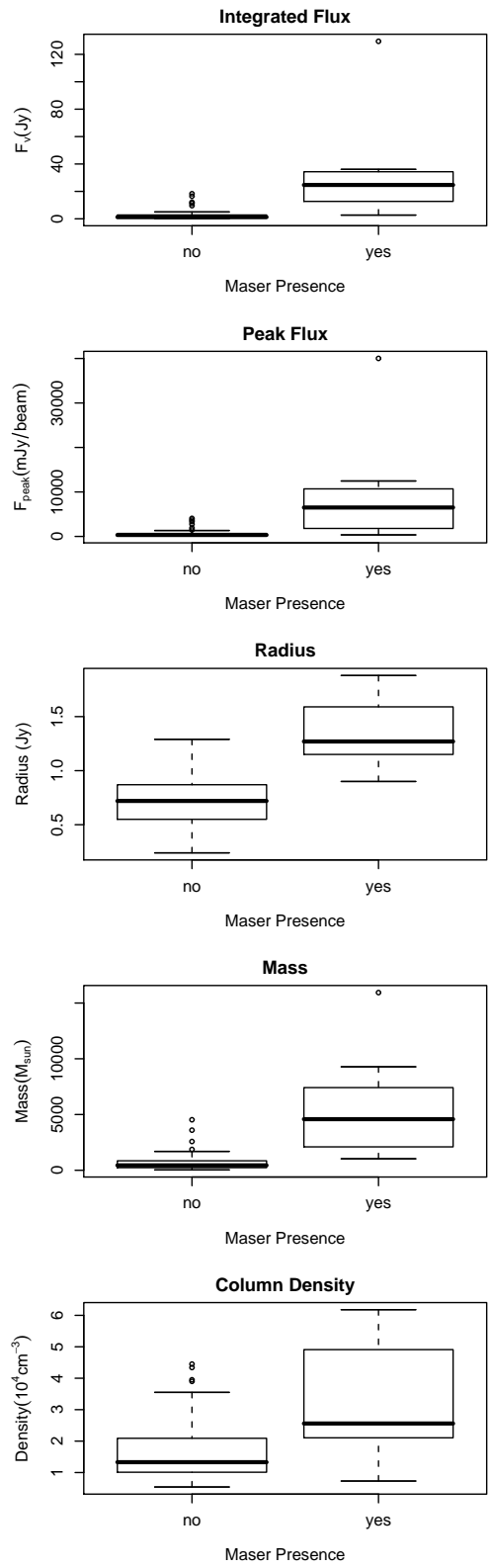,angle=270,scale=0.7}
\caption{Box plots of the 1.2-mm dust clump properties split into
    the categories of yes and no, referring to maser presence and
    absence respectively.}
  \label{fig:dust_box}
\end{figure}

In terms of the 1.2-mm dust clump properties, fits of the single term
addition Binomial model showed an increasing probability that the
presence of a maser was associated with increasing value of all of the
clump properties reported by \citet{Mook04}.  Table~\ref{tab:dust_sta}
gives a summary of the single term addition. This means that any one
of the clump properties may give an indication of the likelihood of
maser presence. There is a significant difference between the clumps with
associated water masers and those without, evident from
Fig.~\ref{fig:dust_box}. Clumps with associated masers are bigger,
denser, more massive and have higher flux densities than clumps where we see no
associated water maser.  The most parsimonious model for predicting
maser presence involved only the radius of the 1.2-mm dust clumps. This equation allows the probability
of maser presence to be predicted knowing only the radius of a 1.2-mm
dust clump. The estimated regression relation is

\begin{eqnarray*}
  \mbox{log}{\frac{p_{i}}{1-p_{i}}}=-11.477+9.174x_{radius},
\end{eqnarray*}
where {\em x}$_{radius}$ is the radius
of the 1.2-mm dust clump in pc. The regression summary of this model is shown in
Table~\ref{tab:dust_regression}.

\begin{table}
  \caption{Analysis of deviance table for the single term models (using the 1.2-mm dust clump properties), showing the AIC and the deviance together with the associated likelihood ratio statistic and p-value for the test of the hypothesis that the stated single model provides no better fit than the null model consisting only of an intercept.}
\begin{tabular}{ccccc}\hline
	{\bf Predictor}& {\bf AIC}  & {\bf Deviance} &  {\bf LRT} & {\bf p-value}\\ \hline
	{\bf none} &   48.130 & 46.130\\
	{\bf F$_{peak}$}  &28.088 &24.088  &22.042& 2.668e--06\\
	{\bf Radius}  &24.330 & 20.330 &25.800 &3.787e--07\\
         {\bf F$_{v}$}   &25.428 &21.428 &24.702 &6.693e--07\\
         {\bf Mass}  &27.467  &   23.467 &22.663 &1.930e--06\\
         {\bf Density} & 39.969 & 35.969 &10.160  &0.001435\\ \hline
         \label{tab:dust_sta}
	\end{tabular}
\end{table}

\begin{table}
	\caption{Summary table for the Binomial regression model, showing for each predictor the estimated coefficient and the standardised z-value and p-value for the test of the hypothesis that $\beta_{i}$=0.}
\begin{tabular}{ccccc}\hline
	{\bf Predictor}& {\bf Estimate} & {\bf Std. Error} & {\bf z value} & {\bf p-value}\\ \hline
	{\bf Intercept} & --11.477 & 3.537 & --3.245 & 0.00118\\
	{\bf Radius} & 9.174 & 3.163 & 2.900 & 0.00373\\ \hline
         \label{tab:dust_regression}
	\end{tabular}
\end{table}

The misclassification rates for the model, given a probability of 0.5
of maser presence within a given clump are again low in predicting the
clumps that have no associated water maser. Of the 73 1.2-mm dust
clumps that fall within our survey region, seven have an associated
water maser, while 66 do not. Our model predicts that 65 of the 66
clumps with no associated water maser will not have an associated
water maser and returns a false-positive result for the remaining
clump. The model predicts that of the seven clumps that have an
associated water maser only four will have an associated water maser
and returns a false-negative for the remaining three clumps.

As this model is only concerned with clump radius it is easy to illustrate its physical implications. By
setting the probability of maser presence to be 0.5, for example, we are able to determine that the corresponding clump radius
is approximately 1.25 pc. This means that all 1.2-mm dust clumps with a radius of 1.25 pc or greater have a probability of 0.5 or more of having an associated water maser.

\section{Conclusions}

Regions within the GMC associated with RCW 106 have been surveyed for 22-GHz water masers. This resulted in the detection of nine water masers (five of these being new detections) and four 22-GHz continuum sources. All of the water masers that we observed have exhibited some level of variability over the 11 month course of these observations. The most extensive temporal variability was observed in water maser G\,333.060--0.488, which showed a variation in peak flux density of a factor of 10 over the observational period. The GMC has previously been searched for 6.7-GHz methanol masers \citep{Simon96} and main-line OH masers \citep{Caswell}. In addition to the nine water masers detected there are four 6.7-GHz methanol masers \citep{Simon96} and three OH masers \citep{Caswell} within the regions surveyed here. All of the three species of masers have sub-arcsecond positional accuracy which allows a comparison of the relative positions of the respective maser species. The water masers that we detect lie along the main axis of star formation within the GMC while the methanol masers are located near the periphery. We find there to be little overlap between the sites of the different maser species, in fact there are only two associations with other maser species. There is one association with a 6.7-GHz maser and one association with an OH maser. 
 
Four of the new water maser detections are associated with GLIMPSE point sources, of similar colours to those associated with detected 6.7-GHz methanol masers. There is a slight bias for the water maser associated sources to be less red. This coupled with the relative positions of the water masers and the 6.7-GHz methanol masers with respect to the main axis of star formation lends support to the hypothesis that 6.7-GHz methanol masers trace an earlier evolutionary phase than water masers.

All of the water masers are associated with a \CO emission peak that we identify or a clump reported by \citet{B06}. Statistical investigation of the \CO \citep{B06} and 1.2-mm dust \citep{Mook04} clumps shows that there is a strong increase in likelihood of water maser detection with increased clump radius, mass, density and brightness. After fitting a Binomial generalized linear model to the maser presence data using the clump properties of \citet{B06} and \citet{Mook04} as predictors we obtained the simplest models with the greatest maser predictive power.  In the case of the 1.2-mm dust clumps our model uses only clump radius to predict the likelihood of the clump having an associated water maser. However the model generated for \CO clumps takes into account radius, integrated flux density of the clump peak and mass. These models have a low misclassification rate and may allow more efficient targeted searches for water masers (where appropriate \CO 
or 1.2-mm data is available) than those previously conducted towards other species of masers or mid-infrared sources. While our survey was of comparatively small scale, we believe that the results are indicative of the likelihood of occurrence of water masers with respect to \CO and 1.2-mm dust clumps.

Observations of \CO clumps and 1.2-mm dust clumps within the G\,333.2--0.6 GMC that do not fall within the regions already surveyed will allow us to
test and refine our predictive models. Additional observations towards
\CO clumps and 1.2-mm dust clumps which are not part of the G\,333.2--0.6 GMC
will allow us to determine models with more highly accurate predictive properties
and lower standard errors. As more data for the G\,333.2-0.6 GMC in other molecular 
line transitions becomes avaliable, analysis similar to what we have undertaken
here will be carried out. The multitude of data avaliable for this GMC 
provides a unique oppourtunity which may provide 
the evidence to support an evolutionary argument.

\section*{Acknowledgements}

We would like to thank the referee Anita Richards for many useful comments that have improved the paper. Thanks also to Robin Wark for her assistance with the ATCA observations. Financial support for this work was provided by the Australian
Research Council. MJH acknowledges support through IRGS grant J0015125 
administered by the University of Tasmania. This research has made use of NASA's Astrophysics
Data System Abstract Service. This research has made use of the NASA/
IPAC Infrared Science Archive, which is operated by the Jet Propulsion
Laboratory, California Institute of Technology, under contract with
the National Aeronautics and Space Administration. This research has
made use of the SIMBAD data base, operated at CDS, Strasbourg,
France. This research has made use of data products from the GLIMPSE
survey, which is a legacy science program of the {\em Spitzer Space
  Telescope}, funded by the National Aeronautics and Space
Administration.  This research has made use of data products from the
{\em Midcourse Space Experiment}.  Processing of the data was funded
by the Ballistic Missile Defence Organization with additional support
from the NASA Office of Space Science.  The research has made use of
the NASA/IPAC Infrared Science Archive which is operated by the Jet
Propulsion Laboratory, California Institude of Technology, under
contract with the National Aeronautics and Space Administration.

\end{document}